\documentclass[11pt,twoside]{article}
\usepackage{asp2010}

\resetcounters

\begin{document}

\title{V496 Scuti: Detection of CO emission and dust shell in a moderately fast Fe II nova}
\author{Ashish Raj,$^1$ N. M. Ashok,$^1$, D. P. K. Banerjee$^1$, U. Munari$^2$, P. Valisa$^2$ and S. Dallaporta$^2$
\affil{$^1$Astronomy and Astrophysics Division, Physical Research Laboratory, Navrangpura Ahmedabad, 380009 Gujarat India}
\affil{$^2$INAF Astronomical Observatory of Padova, 36012 Asiago (VI), Italy}}

\begin{abstract}

We present near-infrared and optical observations of moderately fast FeII-class Nova Scuti 2009 (V496 Sct) covering various
phases; pre-maximum, early decline and nebular, during the first 10 months after its discovery followed by limited observations 
up to 2011 April. In the initial phase the nova spectra show prominent P Cygni profiles and later all the lines are seen in emission. 
The notable feature of the near-IR spectra in the early decline phase is the rare presence of the first overtone bands of carbon monoxide (CO) in emission. 
The IR spectra show clear dust formation in the expanding ejecta at later phase about 150 days after the peak brightness. The presence of lines of
elements with low ionization potentials like Na and Mg in the early IR spectra and the detection of CO bands in emission and the dust formation in 
V496 Sct represents a complete expected sequence in the dust formation in nova ejecta. The light curve shows a slow rise to the maximum and a slow
decline indicating a prolonged mass loss. This is corroborated by the strengthening of P Cygni profiles during the first 30 days. The broad and
single absorption components seen in many lines in the optical spectra at the time of discovery are replaced by two sharper components in the spectra
taken close to the optical maximum brightness. These sharp dips seen in the P Cygni absorption components of Fe II and H I lines during the early
decline phase show increasing outflow velocities. The onset of the nebular phase is evident from the optical spectra in 2010 March. During the 
nebular phase, several emission lines display saddle-like profiles. In the nebular stage, the observed fluxes of [O III] and Hβ lines are used 
to estimate the electron number densities and the mass of the ejecta. The optical spectra show that the nova is evolved in the $P_{fe}A_o$ spectral sequence.
The absolute magnitude and the distance to the nova are estimated to be M$_V$ = −7.0 $\pm$ 0.2 and d = 2.9 $\pm$ 0.3 kpc, respectively.
\end{abstract}

\section{Introduction}

Nova Scuti 2009 (V496 Sct) was discovered by Nishimura on 2009 November 8.370 UT at $V$ = 8.8 (Nakano et al. 2009) on two 10s unfiltered CCD images.
 The low resolution spectra obtained during the period 2009 November 9.73 UT 
to 10.08 UT which showed prominent H${\rm{\alpha}}$ and H${\rm{\beta}}$
 emission lines with P Cygni components, along with the strong Fe II multiplets and O I lines indicating that V496 Scuti is an Fe II class 
nova near maximum light (Munari et al. 2009a, Balam \& Sarty 2009). The optical observations by Munari et al. (2009b) showed
a post-discovery brightening for about 10 days before the onset of fading with maximum brightness $V_{max}$ = 7.07 around 2009 November 18.716 UT. The IR observations
by Rudy et al. (2009) on 2009 November 27.08 UT showed strong first overtone CO emission bands - an extremely short lived feature that is seen in only a few novae.
They also predicted that dust formation in V496 Sct is almost certain. The first result by Raj et al. 2009 showed the continuation of CO emission during the period 2009 December 3.55 UT to 8.55 UT. 
Subsequent observations by Russell et al. (2010) after V496 Sct came out of the solar conjunction showed dust formation on 2010 February 10. 

The IR and optical observations for nova V496 Sct are taken from Mt. Abu IR Observatory of PRL in India, at Asiago Observatory of University of 
Padova and INAF Astronomical Observatory of Padova and Schiaparelli Observatory in Italy, respectively.
\begin{figure}
\centering
\includegraphics[width=4.0in,height=2.0in, clip]{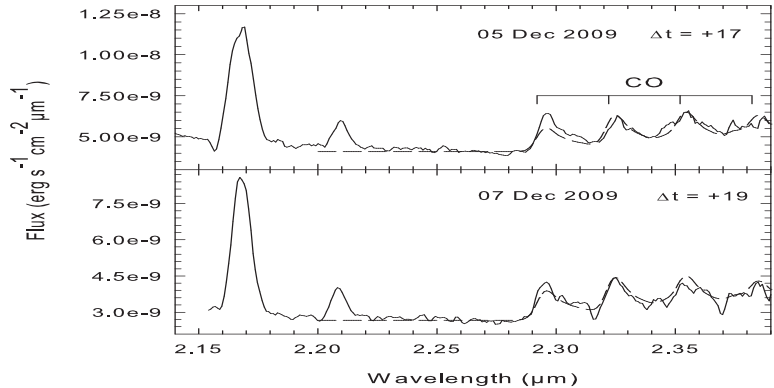}
  \caption[]{ The model fits are shown as dashed lines to the observed first overtone CO bands in V496 Sct for 2009 December 5 and 7. The fits are made for a constant CO mass of 2e-8 M$_\odot$ on both the days while the temperature of the gas T$_{CO}$ is 4000 K and 3600 K respectively. The time from optical maximum are given for each spectrum. }
  \label{fig5}
  \end{figure}

\section{Results}

\subsection{The optical light curve: the pre-maximum rise, outburst luminosity, reddening and distance}
From the optical light curve we estimate $t_2$ to be 59 $\pm$ 5 d (Raj et al. 2012) which makes V496 Sct as one of the moderately fast Fe II class of novae in recent years. 
V496 Sct is one of the large amplitude novae observed in recent years with $\bigtriangleup R$ $\ge$ 13.5 magnitudes (Guido \& Sostero 2009). These observed values of the 
amplitude and $t_2$ for V496 Sct put it above the upper limit in the observed spread of the amplitude versus decline rate plot for classical novae 
presented by Warner (2008, Figure 2.3) which shows $\bigtriangleup V$ = 8 - 11 for $t_{2}$ = 59 days. The height from the galactic plane is estimated 
to be $z$ = 89 $\pm$ 3 pc by using the value of the ditance $d$ = 2.9 $\pm$ 0.3 kpc (Raj et al. 2012) to the nova.
The outburst luminosity of V496 Sct as calculated from M$_V$ is L$_O$ $\sim$ 5.1 $\times$ 10$^4$ L$_\odot$.
A small plateau is also seen in the light curve between 128 and 215 days after the outburst. 

\subsection{Modeling and evolution of the CO emission}

We adopt the model developed in the earlier work on V2615 Oph by Das et al. (2009) to characterize the CO emission. 

\begin{figure*}
\begin{center}$
\begin{array}{cc}
\includegraphics[width=2.4in,height=2.55in, angle=270,clip]{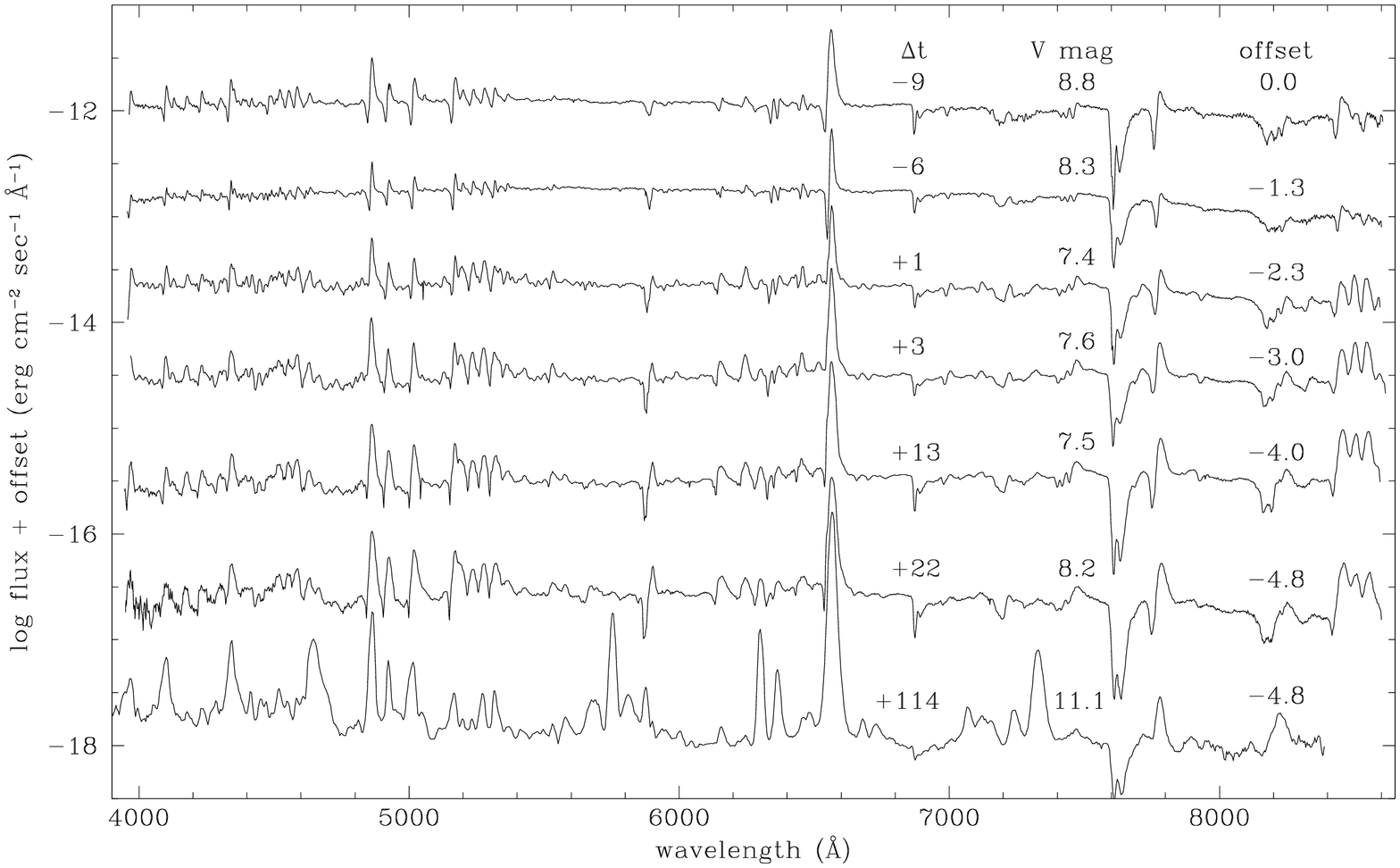}&
\includegraphics[width=2.4in,height=2.55in, angle=270,clip]{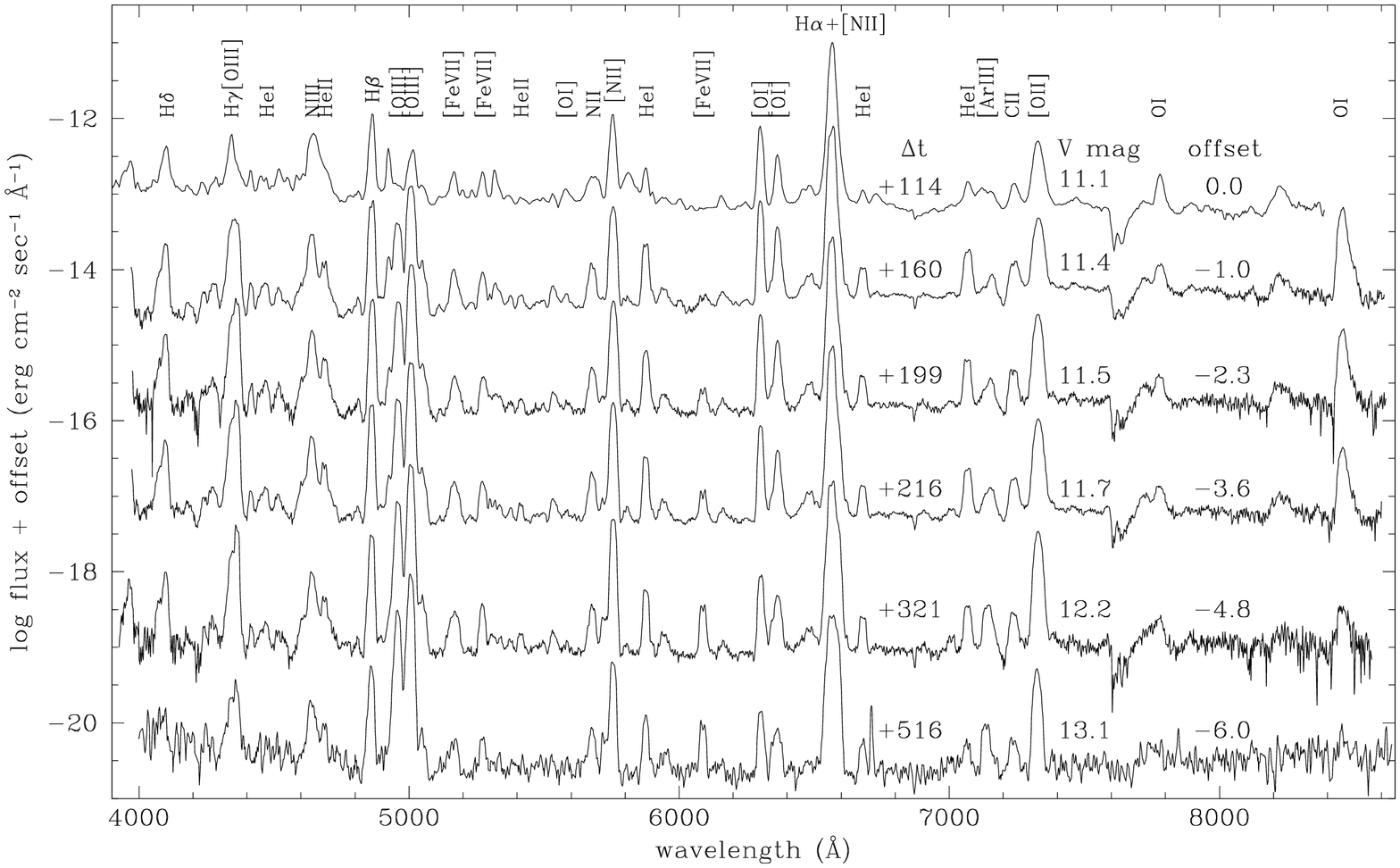}
\end{array}$
\end{center}
  \caption[The $J$ band images (2 $\times$ 2 arcmin) of V496 Sct.]{The low-resolution optical spectra of V496 Sct. The left panel shows the permitted phase ($P_{fe}$) and
right panel shows the auroral phase ($A_o$).}
  \label{ch4_1}
  \end{figure*}

The best fit model spectra to the observed data are obtained by varying the input parameters M$_{CO}$, ${\rm{\alpha}}$, T$_{CO}$ (see Fig. 1). The increase in M$_{CO}$ enhances the 
absolute level of the CO emission while the increase in T$_{CO}$ changes the relative intensities of different vibrational bands in addition to 
changing the absolute level of the emission. The CO emission is assumed to be optically thin. The C I lines at 2.2906 and 2.3130 ${\rm{\mu}}$m and 
Na I lines at 2.3348 and 2.3379 ${\rm{\mu}}$m are also likely to be present in the spectral region covered by the CO emission giving rise to some 
deviations between the best model fit and the observed spectra. The 
typical errors to the formal model fits are $\pm$ 500 K. The model spectra with a reasonably similar range in mass of 
M$_{CO}$ = 1.5$-$2 $\times$ 10$^{-8}$ M$_\odot$ fit the observed spectra. The model calculations also show that the $\nu$ = 2 - 0 bandhead of
$^{13}$CO at 2.3130 ${\rm{\mu}}$m becomes discernibly prominent if the $^{12}$C/$^{13}$C
ratio is $\le$1.5. As this spectral feature is not clearly detected in our observed spectra, we place a lower limit of ∼1.5 for the $^{12}$C/$^{13}$C
ratio.

\subsection {Evolution of the optical spectra and estimation of ejecta mass}

The various phases of the spectral evolution of V496 Sct have been identified using the Tololo classification
 system for novae (Williams et al. (1991); Williams et al. (1994)). The permitted lines of Fe II were the strongest non-Balmer lines in 
the pre-maximum as well as the early decline phase indicating P$_{fe}$ class for the nova. The nova had evolved to the auroral phase A$_{o}$ in 
2010 March as the [N II] 5755 auroral line was the strongest non-Balmer line. We note the absence of [Fe X] 6375 coronal emission line in the 
spectra taken as late as 2011 April 19 (see Fig. 2). Thus the optical spectra show that the nova evolved in the P$_{fe}$A$_{o}$ spectral sequence.
\begin{figure}
\centering 
\includegraphics[width=4.0in,height=1.6in,clip]{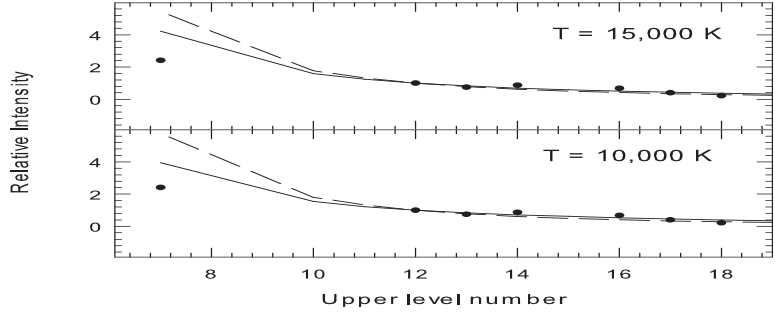}
\caption[The case B analysis of V496 Sct]{ The two panels show the case B analysis for 2009 December 6 for two different temperatures. 
The abscissa is the upper level number of the Brackett series line transition. The line intensities are relative to that of Br 12. The Case B model
 predictions for the line strengths are also shown for a temperature of T = 10$^{4}$ K and electron densities of n$_e$ = 10$^{12}$ cm$^{-3}$
 (dashed line) and 10$^8$ cm$^{-3}$ (solid line).}
  \label{ch4_17}
  \end{figure} 
We have also tried to estimate the ejecta mass by using recombination line analysis of H I lines for 2009 December 6 (Fig. 3). However, we find 
that the strengths of these lines, relative to each other, deviate considerably (specially for Br${\rm{\gamma}}$, Fig. 3) from Case B values for 
2009 December 6 indicating that the lines are optically thick. Hence we are unable to estimate the ejecta mass from recombination analysis. 
From the optical spectra, we estimated the mass of oxygen M$_{O I}$ in the range 1.18$\times$10$^{-5}$-2.28$\times$10$^{-6}$ M$_\odot$. 
The mass of hydrogen m(H) in the ejecta is (6.3 $\pm$ 0.2)$\times$10$^{-5}$M$_\odot$. We obtain M$_{dust}$ = 1-5$\times$10$^{-10}$M$_\odot$ for 2010 April 30 from the best fit value T$_{dust}$ $=$ 1500 $\pm$ 200 K 
(with $\chi$$^2$ minimization) for $d$ = 2.9 kpc.  Hence the gas to dust ratio is found to be M$_{gas}$/M$_{dust}$ $\sim$ 1.3 - 6.3 $\times$10$^{5}$ indicating that a small amount of dust was formed in V496 Sct comparable to 
3 $\times$10$^{5}$ observed in the case of V2362 Cyg by Munari et al. (2008).

\acknowledgements 
The research work at Physical Research Laboratory is funded by the Department of Space, Government of India. We would like to thank
A. Frigo, V. Luppi, L. Buzzi, A. Milani, G. Cherini, A. Maitan, L. Baldinelli (ANS Collaboration).

\end{document}